# Designing generalisation evaluation function through human-machine dialogue

Patrick Taillandier[1], Julien Gaffuri[2]

[1]IRD, UMI UMMISCO 209,
32 avenue Henri Varagnat, 93143 Bondy, France
Email: patrick.taillandier@gmail.com

[2]IFI, MSI, UMI 209,
ngo 42 Ta Quang Buu, Ha Noi, Viet Nam

[3] IGN – COGIT laboratory – Paris-Est University
73 avenue de Paris, 94165 Saint-Mandé cedex, France
Email: julien.gaffuri@ign.fr

## 1. Introduction

A classic approach in automated generalisation consists in formalising generalisation as an optimisation problem: the goal is to find a state of the data that maximises an evaluation function that is supposed to assess the generalisation state of the data, according to the user need (e.g. Wilson et al., 2003). A key issue of this approach concerns the design of this evaluation function. Unfortunately, designing such a function remains a difficult task. Indeed, while the final user of the generalised data can easily describe his/her need in natural language, it is often far more difficult to express his/her expectations in a formal language that can be used by generalisation systems.

In this paper, we propose an approach dedicated to generalisation evaluation functions design. An evaluation function previously designed by a user is improved through a dialogue between the user and a generalisation system. The idea is to collect user preferences by letting him/her compare different generalisation results for a same object.

In Section 2, the context of this work is introduced. Section 3 is devoted to the presentation of our approach. Section 4 describes an experiment carried out for building generalisation and Section 5 concludes.

## 2. Context

### 2.1 Automated evaluation of generalisation results

If many works focus on the generalisation process automation, only a few deal with automatic evaluation of generalisation outcomes. A classic approach consists in evaluating the generalisation quality by means of a set of constraints translating the expectation towards the generalisation (Beard, 1991). The constraint assessment is often represented by a numeric satisfaction value. The overall generalisation is evaluated by aggregating all the constraint satisfaction values. If the computation of individual constraint satisfaction values is often well-managed, the definition of the aggregating function remains complex (Bard, 2004). This paper focuses on this problem.

## 2.2 Design of an evaluation function

The evaluation function design is a complex problem which was studied in various fields. Many works were interested in the definition of these functions for specific problems (Wimmer et al., 2008) but few proposed general approaches for helping optimisation systems users to define it.

A classic approach to solve this problem consists in using supervised machine learning techniques. These techniques consist in inducing a general model from examples labeled by an expert. In this context, it is possible to learn an evaluation function from examples assessed by an expert. This approach was used in several works, like (Wimmer et al., 2008) in the domain of computer vision, and (Clancy et al., 2007) for the learning of cognitive radio. The drawback of this approach is the complexity for experts to quantitatively evaluate the quality of a solution. Indeed, it is sometimes difficult for experts to directly translate the quality of a solution by a numeric value.

## 2.3 Formalisation of the evaluation function design

We assume that a set of constraints, which assessment is represented by a numeric satisfaction value, is defined. The higher the assessment value, the more satisfied the constraint is, thus better the generalisation is. We propose to formulate the aggregation function by a weighted means balanced by a power.

Let $C$ be the constraint set considered, $w_i$ the weight associated to a constraint $i$, $Val_i(gen)$, the assessment value of the constraint $i$ for the generalisation $gen$, and $p$, an integer higher or equal to 1. The evaluation function is defined as follows:

$$quality(gen) = \left[ \frac{1}{\sum_{i \in C} w_i^p} \cdot \sum_{i \in C} (w_i^p \cdot Val_i(gen)^p) \right]^{\frac{1}{p}}$$

The role of $p$ is to control the relative weight of the most satisfied constraints over the less satisfied ones: the higher is $p$, the more satisfied constraints are taken into account in the overall quality of the generalisation.

## 3. Proposed approach

### 3.1. General approach

We propose an approach to design the generalisation evaluation function based on the presentation of comparisons between generalisations to the user and the learning of an evaluation function from the collected preference data. This approach is close to the one proposed by (Hubert & Ruas, 2003) concerning the parameterisation of the generalisation process. However, a difference is that the user will not just select his/her preferred generalisation among a set, but the user will compare these generalisations. Our approach is composed of 3 steps that are described in the next sections.

### 3.2. Initialisation of the comparison set

The first step concerns the generation of the comparisons, which will be shown to the user to capture his/her needs. A comparison is a pair of different generalisations of the same object. In order to build the comparison set, some geographic objects to generalise are selected. Two different generalisations of these objects are then computed and stored in the comparison set.

### 3.3 Capture of the user preferences

The second step concerns capture of the user preferences: comparisons are successively presented to the user, who gives his/her preference for each of them. This sequence is reiterated until a specific number of comparisons have been presented to the user.

For each comparison between two generalisations *A* and *B*, the user can choose:
- Generalisation *A/B* is far better than Generalisation *B/A*
- Generalisation *A/B* is better than Generalisation *B/A*
- Generalisation *A/B* is slightly better than Generalisation *B/A*
- Generalisation *A* and Generalisation *B* are equivalent

Figure 1 presents the comparison interface of the developed prototype.

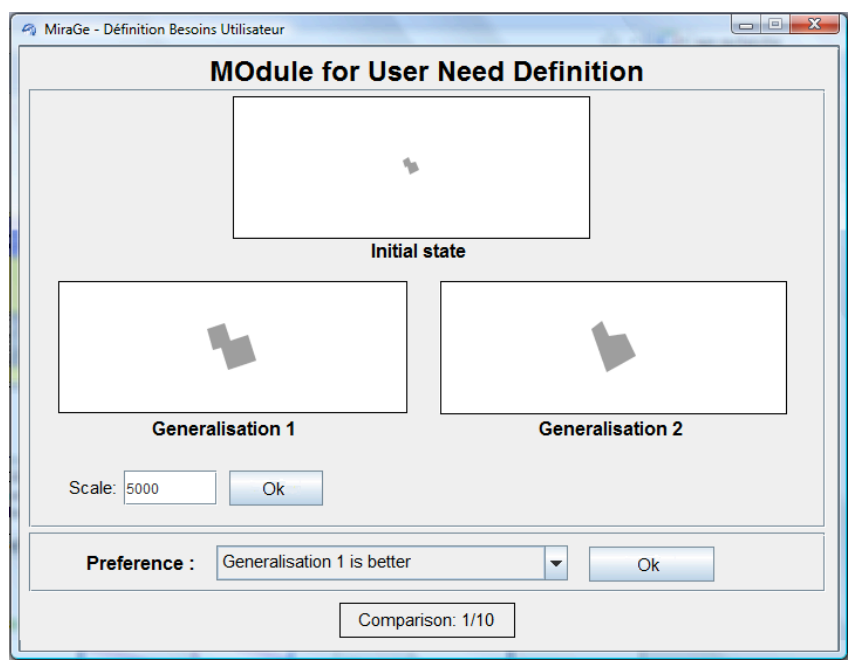

Fig. 1. The comparison interface

### 3.4. Evaluation function definition

The last step consists in learning an evaluation function from the captured user preferences: the parameter values (i.e. the constraint weights $w_i$ and the power $p$) that best fits the preferences given by the user during the previous step are computed. We propose to formulate this problem as a minimisation problem. We define a global error function that represents the inadequacy between an evaluation function (and thus the parameter values assignment) and the user preferences. Our goal is to find the parameter values that minimise the global error function.

Let $f_{eval}(gen)$ be the current evaluation function that evaluates the quality of a generalisation *gen*; $c_{gen1,gen2}$ a comparison between two generalisations, $gen_1$ and $gen_2$; $p_c$ the user preference for the comparison *c*. We define the function $comp(c, f_{eval}, p_c)$ that determines for a comparison *c* if the user preference $p_c$ is compatible with the evaluation function $f_{eval}$, i.e. if the preference is consistent with the quality order obtained by applying the evaluation function. $comp(c, f_{eval}, p_c)$ is computed as follows:

$$comp(c, f_{eval}, p_c) = \begin{cases} 0 \text{ if } \begin{cases} p_c = gen_1 \text{ is far better than } gen_2 & \text{and} & f_{eval}(gen_1) - f_{eval}(gen_2) \geq Val_{min}^{FB} \\ \text{or } p_c = gen_2 \text{ is far better than } gen_1 & \text{and} & f_{eval}(gen_2) - f_{eval}(gen_1) \geq Val_{min}^{FB} \\ \text{or } p_c = gen_1 \text{ is better than } gen_2 & \text{and} & Val_{max}^{B} \geq f_{eval}(gen_1) - f_{eval}(gen_2) \geq Val_{min}^{B} \\ \text{or } p_c = gen_2 \text{ is better than } gen_1 & \text{and} & Val_{max}^{B} \geq f_{eval}(gen_2) - f_{eval}(gen_1) \geq Val_{min}^{B} \\ \text{or } p_c = gen_1 \text{ is slightly better than } gen_2 & \text{and} & Val_{max}^{SB} \geq f_{eval}(gen_1) - f_{eval}(gen_2) \geq Val_{min}^{SB} \\ \text{or } p_c = gen_2 \text{ is slightly better than } gen_1 & \text{and} & Val_{max}^{SB} \geq f_{eval}(gen_2) - f_{eval}(gen_1) \geq Val_{min}^{SB} \\ \text{or } p_c = gen_1 \text{ and } gen_2 \text{ are equivalent} & \text{and} & |f_{eval}(gen_1) - f_{eval}(gen_2)| \leq Val^{Eq} \end{cases} \\ 1 \quad otherwise \end{cases}$$

This formula introduces the parameters $Val_{min}^{FB}$, $Val_{min}^{B}$, $Val_{max}^{B}$, $Val_{min}^{SB}$, $Val_{max}^{SB}$ and $Val^{eq}$ that confer a fuzzy aspect to the notion of compatibility.

The global error function proposed corresponds to the percentage of comparisons of the comparison sample *C* that are incompatible with the evaluation function $f_{eval}$:

$$Error(f_{obj}, C) = \frac{100}{|C|} \cdot \sum_{c \in C} comp(c, f_{eval}, p_c)$$

Parameter values that minimise *Error(f<sub>obj</sub>, Comp)* have to be found. In order to facilitate the search process, we propose to use an evaluation function initially defined by an expert. Indeed, we make the hypothesis that, most of the time, experts –that usually have a good command of the generalisation system-- can design a good generic evaluation function, which can be adapted for some more specific needs. In consequence, we propose to use a local search algorithm and more particularly a tabu search (Glover, 1989). The principle of this kind of algorithm is to start with an initial solution and to attempt to improve it by exploring its neighbourhood. These algorithms are usually very effective for this kind of search problem. Local search algorithms require the definition of the notions of 'solution' and 'solution neighbourhood'. For our problem, a solution is a parameters assignment (weights $w_i$, and power *p*). We define the neighbourhood of a solution as the set of solutions for which only one parameter has its value changed.

## 4. Case study: evaluation of building generalisation

### 4.1. Context

Our experiment use a generalisation system based on the AGENT model (Barrault et al., 2001). In this model, the quality of the generalisation is evaluated by a set of constraints. The AGENT model has been the core of numerous research works and is used for map production in several mapping agencies. However, the question of the constraint weight assignment is still asked (Bard, 2004).

We propose to experiment our method for building generalisation for a traditional 1:25000 scale topographic map. Five constraints are used. The input data are taken from the BDTopo®, a one meter resolution topographic database.

The initial evaluation function was designed by an expert of the AGENT model.

We defined two set of 100 different comparisons each (the *learning* and the *testing* set). The *learning* set is used to learn the evaluation function, the *testing* to assess the quality of the initial and learnt evaluation functions.

### 4.2. Results and discussion

Table 1 presents the results on the two comparison sets. It shows for each evaluation function and comparison sets the global error (c.f. Section 3.4).

|  | Global error | |
| --- | --- | --- |
|  | *Initial function* | *Learnt function* |
| **Learning set** | 44.1% | 27.4% |
| **Test set** | 40.1% | 29.0% |

Table 1. Results

These results reveal that the learnt function has allowed an improvement of the global error: for both learning and testing sets, the global errors of the initial function are higher than for the learnt function. However, the quality improvement after the use of the method is only of 11% for the test set. An explanation is the lack of constraints (for example, an orientation constraint). For example, when a comparison composed of two building generalisations, which differ only in term of orientation is shown, the user always prefers the one whose orientation is close to the building initial orientation. Because there is no orientation constraint taken into account into the evaluation function, the difference of the two generalisations can not be measured by the system, and the reason of the different assessment by the user remains ignored. In this context, our approach, through an examination of the incompatible comparisons, can help to determine some important missing constraints and identify faulty ones.

## 5. Conclusion

In this paper, we presented an approach dedicated to the definition of a generalisation evaluation function. We proposed a method based on a human-machine dialogue and the capture of user preferences on generalisation samples. An experiment carried out in the domain of cartographic generalisation showed how our approach can help users to define better evaluation functions.

This work is at its beginning. In the near future, we plan to carry out more experiments, in particular to study the impact of the initial evaluation function on the results.

Our long-term purpose is to provide a method to learn the user preferences concerning all objects and group of objects contained in its data. The last stage of this research would be to automatically learn a user final evaluation method for a complete map piece. Such a system would be able to make an automatic interview of the user, allowing him to give his specific requirements for all characteristics of the map.

## References

Bard S, 2004, Quality Assessment of Cartographic Generalisation, *Transactions in GIS*, 8, 63-81.

Barrault M, Regnauld N, Duchêne C, Haire K, Baeijs C, Demazeau Y, Hardy P, Mackaness W, Ruas A and Weibel R, 2001, Integrating multi-agent, object-oriented, and algorithmic techniques for improved automated map generalization. *ICC*, 2110-2116

Beard K, 1991, Constraints on rule formation. In Buttenfield, B. & Mcmaster, R. (eds.) Map generalisation: making rules for knowledge representation, 121-135.

Clancy C, Hecker J, Stuntebeck E and O'Shea T, 2007, Applications of Machine Learning to Cognitive Radio Networks, *Wireless Communications*, vol. 14, no. 4, 47-52

Glover F, 1989, Tabu search. Journal on Computing.

Hubert F and Ruas A, 2003, A method based on samples to capture user needs for generalisation, *workshop on progress in automated map generalisation*, Paris.

Wilson ID, Ware JM and Ware JA, 2003, Reducing graphic conflict in scale reduced maps using a genetic algorithm. *Workshop on progress in automated map generalisation*, Paris.

Wimmer M, Stulp F, Pietzsch S and Radig B, 2008, Learning local objective functions for robust face model fitting. *IEEE Transactions on Pattern Analysis andMachine Intelligence*, 30(8).